\documentstyle[12pt,epsfig,cite]{article}

\newlength{\dinwidth}
\newlength{\dinmargin}
\def\titlepage{\clearpage%
\setcounter{footnote}{0}\setcounter{page}{1}%
\thispagestyle{empty}\pagestyle{plain}\pagenumbering{arabic}%
\kern1mm\begin{center}
\end{center}
\vskip3mm\normalsize}
\def\docnum#1{\hbox to \hsize{\hskip123mm\hbox{#1}\hss}}
\def\date#1{\edef\@temp{#1}\ifx\@temp\@empty\def\@temp{\today}\fi
\hbox to \hsize{\hskip123mm\hbox{\@temp}\hss}}
\def\title#1{\vskip 0.8in plus 2in\begin{center}%
{\Large\bf#1\par}\vskip1.5em\end{center}\vskip 1in}
\def\@makefnmark{\hbox{$^{\@thefnmark)}$}}
\def\author#1{
\setcounter{footnote}{0}\def\@currentlabel{}%
\begingroup\def\thefootnote{\arabic{footnote}}
\def\@makefnmark{\hbox{$^{\@thefnmark)}$}}
\global\@topnum\z@ \large\begin{center}{\lineskip.5em
\begin{tabular}[t]{c}#1\end{tabular}\par}
\end{center}\par\vskip1.5em\@thanks\endgroup}

\def\abstract{\vskip0.8in plus 3in\begin{center}{\large\bf
Abstract}\end{center}\quotation}

 \newcommand{\bci} {{\bf Q}}
 \newcommand{\bfi} {{\bf \phi}}
 \newcommand{\bcii} {{\bf q}_i}

\begin{document}

\begin{titlepage}
\flushright{DFF 233/10/1995}
\flushright{DFTT 64/95}
\flushright{MPI-PhT/95-105}
\flushright{October 1995}
\title{Multiplicity distributions in a thermodynamical model of hadron
production in $e^+e^-$ collisions
\footnote[1]{Work supported in part by M.U.R.S.T. under grant 1994}}
\vspace{-2.0cm}
\centerline{\large F. Becattini}
\centerline{\it{Universit\`a di Firenze and INFN Sezione di Firenze}}
\centerline{\it{Largo E. Fermi 2, I-50125 Firenze}}
\centerline{e-mail: becattini@vaxfi.fi.infn.it}
\vspace*{0.5cm}
\centerline{\large A. Giovannini}
\centerline{\it{ Universit\`a di Torino and INFN Sezione di Torino }}
\centerline{\it{ Via P. Giuria 1, I-10125 Torino }}
\centerline{e-mail: giovannini@to.infn.it}
\vspace*{0.5cm}
\centerline{\large S. Lupia}
\centerline{\it {Max-Planck-Institut f\"ur Physik,
Werner-Heisenberg-Institut}}
\centerline{\it {F\"ohringer Ring 6, D-80805 M\"unchen }}
\centerline{e-mail: lupia@mppmu.mpg.de}

\begin{abstract}
Predictions of a thermodynamical model of hadron production for
multiplicity distributions in $e^+e^-$ annihilations at
LEP and PEP-PETRA centre of mass energies are shown. The production
process is described as a two-step process in which primary hadrons
emitted from the thermal source decay into final observable particles.
The final charged tracks multiplicity distributions turn out
to be of Negative Binomial type and are in quite good agreement
with experimental
observations. The average number of clans calculated from fitted
Negative Binomial
coincides with the average number of primary hadrons predicted
by the thermodynamical model, suggesting that clans should be
identified with primary hadrons.
\end{abstract}

\end{titlepage}

\section{Introduction}

Multiplicity Distributions (MDs) are fundamental observables in
multiparticle  production in $e^+e^-$ collisions. Theoretical
investigations of their properties, in the QCD framework, are
based on a comparison of approximate calculations at
parton level with experimentally observed final particles
distributions via the
assumption of the Local Parton Hadron Duality \cite{lphd}
or its generalization \cite{glphd}.
In this context the important role of resonances and particles
decay is
not explicitly taken into account. This fact is particularly
unsatisfactory because
a large fraction of final particles are indeed decay products
of heavier ones.
A second approach is based on Monte Carlo models, like
JETSET \cite{jet} and
HERWIG \cite {her}, which account for the decay chain
following the perturbative phase and hadronization.
However, predictions of these models are obtained by
using hadronization schemes requiring a large number
of free parameters in order to reproduce experimental
data.\\
In this paper we discuss the MDs obtained in a thermodynamical
model of
multiparticle production in $e^+e^-$ annihilations, introduced
by one of the authors \cite{td1},
which successfully reproduces the production rates of the
various hadrons species
both at PEP-PETRA and LEP centre of mass energies.
The model is based on the identification of jets with thermalized
hadron gas phases after the hadronization of primary quarks has
taken place.
As a first approximation it is assumed that only two phases with
opposite momenta
are generated in an $e^+e^- \rightarrow q \bar q$ event,
namely multi-jets events are neglected. The observed multi-particle
production in hadronic events is the result of a two-step process:
in the first one some primary hadrons (particles and resonances)
emerge directly from the
thermalized phases after having decoupled. These primary hadrons
then decay
according to known decay modes and branching ratios, giving rise
to observable
particles in the detector. Only three parameters describe those
phases at the decoupling time: the temperature $T$, the volume
$V$ and a
strangeness chemical equilibrium suppression parameter $\gamma_s$.
They are determined by fitting the calculated average
production rates of each hadron to those measured both at
LEP ($\sqrt s \simeq 91$ GeV)
and at PEP-PETRA ($\sqrt s = 29 \div 35$ GeV). A major role
in the determination of the
hadron rates is played by the conservation laws: the quantum
numbers of the jet
are related to those of the primary quark from which the jet
itself originated.
Accordingly, it is assumed that each jet keeps the charm and
beauty of the parent
quark while non-vanishing baryon number and strangeness are allowed
provided that the baryon number and strangeness of the whole
system are zero.

\section{Thermal fluctuations and multiplicity distributions}

Consider a hadron gas at temperature $T$ and volume $V$.
If $T \approx {\cal O}(100)$ MeV, i.e. less than the mass
of all hadrons but pions, the simple Boltzmann statistics
holds for all hadrons except pions, which indeed obey Bose
statistics \cite{td1}.
Therefore, if the quantum numbers of the gas are not fixed,
each species
of hadron fluctuates independently according to a Poisson
distribution, whereas
pions fluctuate according to a different distribution that
we define as the
$\pi$-distribution $f_\pi$ and we derive in Appendix A.
On the other hand, if the flavour quantum
numbers of the gas are fixed, conservation laws generate
correlations between
different hadron species, so that fluctuations are no
longer independent.\\
Therefore, the probability to observe $n_1$ $\pi^+$,
$n_2$ $\pi^0$, $n_3$ $\pi^-$, $n_4$
hadrons of kind 4,..., $n_K$ hadrons of kind $K$ in a hadron
gas system is given by:

 \begin{equation}
     P(n_1,\ldots,n_K) = \frac {1}{Z({\bf Q})} \,
     f_{\pi}(n_1,n_2,n_3) \prod_{i=4}^K \,
   \frac {z_i^{n_i}}{n_i!} e^{-z_i} \,\,
   \delta_{ {\bf Q}, \sum_{i=1}^K n_i {\bf q}_i } \; ,
 \end{equation}
where

\begin{equation}
  z_i = (2J_i+1) \, \frac {V}{(2\pi)^3} \int d^3 p \;
  e^{-\frac {\sqrt{p^2+m^2_i}}{T}} \; ,
\end{equation}
${\bf Q}= (N,S,C,B)$ is a four dimensional vector with
integer components
representing baryon number, strangeness, charm and beauty of the gas;
${\bf q}_i$, $J_i$, $m_i$ are the quantum numbers vector,
the spin and the mass of the
$i^{th}$ particle and $f_{\pi}$ is the $\pi$-distribution. The factor
$\delta_{ {\bf Q}, \sum_{i=1}^K n_i {\bf q}_i }$ accounts
for the conservation of
the mentioned quantum numbers. $Z({\bf Q})$ is the partition
function of the system, namely:

\begin{equation}
  Z({\bf Q}) = \sum_{n_1=0}^\infty \ldots \sum_{n_K=0}^\infty
  f_{\pi}(n_1,n_2,n_3)
  \prod_{i=4}^K \, \frac {z_i^{n_i}}{n_i!} e^{-z_i} \,
  \delta_{ {\bf Q}, \sum_i n_i {\bf q}_i } \; .
\end{equation}
With the substitution

\begin{equation}
   \delta_{ {\bf Q}, \sum_i n_i {\bf q}_i } =
   \frac{1}{(2\pi)^4} \int \,
     d^4 \phi \,\, \exp{ \{ i({\bf Q}- \sum_i n_i
     {\bf q}_i)\cdot \phi) \}} \; ,
\end{equation}
where $\phi=(\phi_1,\phi_2,\phi_3,\phi_4)$, Eq. (3) becomes:

\begin{equation}
     Z(\bci) = \frac{F_\pi}{(2\pi)^4} \int\, d^4 \phi \,\,
     e^{i \bci \cdot \bfi }
     \exp \, \{ {\sum_{i} z_i e^{-i \bcii \cdot \bfi }}\} \; ,
\end{equation}
which is the expression obtained for the hadron gas partition
function in \cite{td1} starting
from the general formulae of partition functions of
thermodynamical systems with
internal symmetry \cite{zf}. The function $F_\pi$ turns out to be:

\begin{equation}
     F_\pi = \exp \, \,\{- \sum_{i=1}^3 \frac {V}{(2\pi)^3}
     \int d^3 p\,\,\log \,
     (1-e^{-\frac {\sqrt{p^2+m^2_i}}{T}})\} \; ;
\end{equation}
where the sum runs over the three pion states.\\
 From Eq. (1) one can build up the expression of
 the joint probability to
observe a $K$-uple ${\bf n}=(n_1,\ldots,n_K)$ of numbers of
primary hadrons in
the first jet and a $K$-uple ${\bf m}=(m_1,\ldots,m_K)$ in
the second jet
%

\begin{equation}
     P({\bf n},{\bf m}) = \frac {1}{\hat Z} \,
     f_{\pi}(n_1,n_2,n_3) \,
     f_{\pi}(m_1,m_2,m_3) \prod_{i=4}^K \,
     \frac {z_i^{n_i}}{n_i!} e^{-z_i}
     \prod_{i=4}^K \, \frac {z_i^{m_i}}{m_i!} e^{-z_i} \,\,
     \delta_{ 0, \sum_i n_i {\bf q}_i + m_i {\bf q}_i} \; ,
\end{equation}
with ${\hat Z}$ given by

\begin{eqnarray}
  {\hat Z} & = & \sum_{n_1=0}^\infty \ldots \sum_{n_K=0}^\infty
  \sum_{m_1=0}^\infty \ldots \sum_{m_K=0}^\infty
  f_{\pi}(n_1,n_2,n_3)
     f_{\pi}(m_1,m_2,m_3) \cdot \nonumber \\
     & & \cdot \prod_{i=4}^K \, \frac {z_i^{n_i}}{n_i!} e^{-z_i}
     \prod_{i=4}^K \, \frac {z_i^{m_i}}{m_i!} e^{-z_i} \,\,
     \delta_{ 0, \sum_i n_i {\bf q}_i + m_i {\bf q}_i} \; .
\end{eqnarray}
It should be pointed out that the partition function of the
two-jet system is further modified by other constraints:
\begin{description}
\item{-} net baryon number per jet $\mid N \mid \,\, < 2$
\item{-} net strangeness per jet $\mid S \mid \,\, < 3$
\item{-} $S \neq 0$ in $e^+e^- \rightarrow s \bar s$ events
\item{-} net charm per jet $\mid C \mid \, =1$ in $e^+e^-
\rightarrow c \bar c$,
 $\mid C \mid \,=0$ otherwise
\item{-} net beauty per jet $\mid B \mid \, =1$ in $e^+e^-
\rightarrow b \bar b$ events,
 $\mid B \mid \,=0$ otherwise
\end{description}
These constraints can be easily implemented by multiplying the
right-hand side of Eq. (7) by additional suitable $\delta$'s
and by recalculating
accordingly the $\hat Z$ in the Eq. (8). Furthermore, the
probability of production
of primary hadrons containing $n$ strange valence quarks
should be multiplied by
a suppression factor $\gamma_s^n$.\\
The Eq. (7) suggests a simple procedure for calculating
the charged tracks
MDs in $e^+e^- \rightarrow q \bar q$
events with a Monte-Carlo method.
The first step of the procedure is to pick up
two $K$-uples $((n_1,\ldots,n_K),(m_1,\ldots,m_K))$ randomly
according to Poisson or $\pi$-distribution and to accept the
event only if the
conditions imposed by the $\delta$'s in Eq. (7) and by the
other requirements
are fulfilled. The second step is to perform all decays of
the generated primary
hadrons according to known
branching ratios, until $\pi$, $K$, $K_L^0$, $\mu$ or
stable particles are reached, in order to match the MDs
measured by $e^+e^-$ colliders experiments, including all
decay products of particles with  $c\tau < 10$ cm.\\

\section{Results}

We calculated the multiplicity distributions by the Monte-Carlo
method described in the previous section at LEP and PEP-PETRA
centre of mass energies.
We used the same values of the parameters $T$,$V$ and $\gamma_s$
determined in \cite{td2} (see Table 1) by fitting the calculated
average production rates of various hadron species to the
measured ones \cite{dean}.
Two additional parameters are introduced in order to take
into account the
possibility of different values of $V$ as a function of
the primary quark
mass, i.e. two dimensionless variables
$x_c, x_b \in [0,1]$ such that $x_c V$ is
the volume in $e^+ e^- \rightarrow
c \bar c$ and $x_b V$ in $e^+ e^- \rightarrow b \bar b$
events, whereas
$V(1-x_c R_c - x_b R_b)/(R_u+R_d+R_s)$ is the volume
for $e^+ e^- \rightarrow q
\bar q$ where $q$ is a light quark and
$R_q = \sigma(e^+ e^- \rightarrow q \bar q)
/\sigma(e^+ e^- \rightarrow hadrons)$.
The factor $x_c$ ($x_b$) has been determined
constraining the difference $\delta_{cq}$
($\delta_{bq}$) between the average charged
multiplicity in $e^+ e^- \rightarrow c \bar c$
($e^+ e^- \rightarrow b \bar b$) and the average
charged multiplicity in $e^+ e^- \rightarrow q \bar q$,
with $q=u, d, s$ for LEP,
$q=u, d, s, c, b$ for PEP-PETRA to be equal to the measured
ones \cite{bclep,bclow}.
The measured values of $\delta_{cq}$ and $\delta_{bq}$ have
been averaged according to the
procedure described in \cite{schm}.\\
We generated 10000 events for each centre of mass energy
in order to get the
MDs of primary hadrons and of charged tracks both in single
hemisphere (i.e. single
jet according to the thermodynamical model) and in
full phase space.\\
The calculated MDs are fitted with a Negative Binomial
(NB) distribution with
Maximum Likelihood method (see Fig. 1a, 1b and 2a, 2b).
Average charged tracks multiplicity $\overline{n}$ and
parameter $k$ (linked to dispersion $D$ by the relation
$D^2=\overline{n}+\overline{n}^2/k$)
obtained from the fit are quoted in Table 2 together
with corresponding values
from analogous NB fits to experimental data \cite{del1,tasso}.
Two kinds of error affect the determination of the NB parameters:
the first one is the fit error, due to the limited statistics
of 10000 events; the
second one is related to the uncertainty
on the parameters of the thermodynamical
model $T$, $V$, $\gamma_s$, $x_c$, $x_b$
used as input in the event generation.
This latter error, a systematic one, has been
estimated by varying the values of parameters
by their error, as quoted in Table 1,
generating a new sample of events and repeating the NB fit.
The covariance matrix $M$ of $\overline{n}$ and $k$ has been
determined with the usual formula
\begin{equation}
M=J C J^{T}
\end{equation}
where $C$ is the covariance matrix of $T$, $V$, $\gamma_s$,
$x_c$, $x_b$ and $J$ is
the jacobian matrix relating $(\overline{n}, k)$ to
$(T,V,\gamma_s,x_c,x_b)$.\\
The obtained systematic errors dominate over statistical
ones, as shown in
Table 2. Since the $\chi^2$ test on the NB fit consistency
is also affected by those
systematic effects, we estimated the uncertainty on $\chi^2$
according to
the same procedure used for $\overline{n}$ and $k$.
Taking this error into account, the $\chi^2$ test
indicates a clear compatibility
with the NB distribution within the uncertainty of
the input parameters.\\
The agreement between the parameters obtained by the
model and those obtained from NB fits
to experimental data is remarkable (see Table 2).
However, discrepancies arise in
the direct comparison between calculated distributions
and measured ones \cite{del2,tasso}, as shown in Fig. 3.
The overestimation of the probability distribution in the
low-multiplicity bins at PETRA energies and the consequent
increase of the dispersion $D$, can be
attributed to violations of the statistical framework of
the canonical ensemble at low centre of mass energies.
It is expected indeed that
the probability of producing a low number of primary
hadrons (for instance 0 or 1)
should be strongly suppressed in comparison to the
predictions of the canonical
distribution, due to the exact energy conservation.
This deviation from the
canonical behaviour is much less visible at LEP energies,
where the average
number of primary hadrons is higher.\\
Another discrepancy is related to the so-called
"shoulder effect": the experimental
charged tracks MD deviates from the NB distribution
due to a shoulder structure
in the high multiplicity region, which is interpreted
as the effect of the superposition of
low-multiplicities 2-jets events with multi-jets events
yielding higher multiplicities.
On the other hand, it has been shown that good NB fits
are obtained with MDs
in selected 2, 3 or 4 jets event samples as well as with
individual jets MDs \cite{del3}.
Since the thermodynamical model is based on a 2-jet scheme
and neglects multi-jets
events, an analogous disagreement with data is expected,
while the agreement with
NB confirms its validity in reproducing MDs in 2-jets events.
It should be mentioned also that a comparison between our
predictions and a 2-jet events data sample is not possible
because the parameters of the model have been
tuned to reproduce the production rates of the
hadron species in the overall
$e^+e^- \rightarrow q \bar q$ data sample.

\section{Comparison with the clan model}

The clan model has been introduced in order to interpret
the wide occurrence of
the NB regularity of charged tracks MDs in all high energy
reactions \cite{nb}. This model is based on the assumption
that initial independent,
i.e. Poissonian, production of primary objects called clans,
is followed in
the second step by their decay into final particles according
to a logarithmic
distribution. The average number of clans is determined in
terms of the $\overline{n}$
and $k$ parameters of the observed Negative Binomial
distribution according  to the formula:
\begin{equation}
\overline{N}_c=k \log \, (1+\overline{n}/k) \; .
\end{equation}
Since the thermodynamical model also contains a two step structure
of the hadron production process consisting in an initial generation
of a number of primary
hadrons from the thermal source which then decay into final
observable particles,
one can ask whether clans can be identified with the primary
hadrons of the
thermodynamical model. The average number of primaries, in
this case, is determined
independently by using the parameters of the thermodynamical
model determined as
mentioned in Sect. 3.\\
Results of this comparative analisys are summarized in Table 3.
The agreement
between the average number of primaries as resulting from the
two models is striking.
It should be noticed that in the clan model primaries are
distributed according
to an exact Poisson distribution, while in the thermodynamical
model the distribution of
primaries can be fitted by a NB distribution with a quite large
$k$ value, which can
be well approximated by a Poisson distribution (see Fig. 1c, 2c).
This small deviation from an exact Poissonian behaviour is a
consequence
of the presence of conservation laws as specified in the
$\delta$'s of Eq. 8, and
of correlations contained in the $\pi$-distribution for
primary pions. It should
be noticed, however, that the $\pi$-distribution, with
the actual values of $T$ and $V$,
is very close to a Poisson distribution (see Table 4).

\section{Conclusions}

The thermodynamical model of hadron production in $e^+e^-$
collisions has been applied to the study of the charged
tracks multiplicity distributions.
Predictions of the model reproduce the gross features
of experimental data
both at LEP and PEP-PETRA centre of mass energies.
The observed discrepancy with experimental measured MD at
LEP energy is due to the presence of multi-jets events
which are neglected in the model (shoulder effect).
A natural violation of the canonical distribution,
expected at low energy, is responsible
for the observed deviation from the experimental measured
MD at PETRA in the low  multiplicity region.
Also, the predicted charged tracks MDs are shown to be in
very good agreement with the
long-standing observed NB regularity and both at LEP and
PEP-PETRA centre of mass energies.
The primary hadrons MD is also well fitted by a NB
distribution with a quite large
value of $k$ parameter, close to the Poissonian limit.
The average number of primary
hadrons, independently determined in the thermodynamical model, is
approximately equal to the average number of clans
calculated from the parameters
of the NB fit to the final charged tracks MDs. This
identity suggests identifying
clans with primary hadrons and confirms the intuition
that the occurrence of Negative
Binomial distribution is deeply related to a two-step
process \cite{nb}.\\
These results confirm previous successful predictions
of the thermodynamical model,
opening a new perspective in the understanding of the
hadronization process at high
energy. An extension of this model to include multi-jets
events is going to be pursued
in order to reproduce also fine features of experimental
multiplicity distributions.

\newpage

\section*{Appendix}
\appendix
\section{Determination of the $\pi$-distribution}

In a boson gas at temperature $T$ and volume $V$ the
probability to observe $n_1$
particles in the kinematical state 1 of energy $\varepsilon_1$,
$n_2$ particles
in the kinematical state $2$ of energy $\varepsilon_2$, ... is:

\begin{equation}
 P(n_1,\ldots,n_M) = \prod_i (1-e^{-\varepsilon_i/T}) \,
 e^{-n_i\varepsilon_i/T} \; .
\end{equation}
The probability that the overall population is $N$
can be written as:

\begin{equation}
 P(N) = \sum_{n_1=0}^\infty \ldots \sum_{n_M=0}^\infty
 \prod_i (1-e^{-\varepsilon_i/T}) \, e^{-n_i\varepsilon_i/T}
 \delta_{N,\sum_i n_i} \; .
\end{equation}
Since:

\begin{equation}
 \delta_{N,\sum_i n_i} = \frac{1}{2\pi} \int_0^{2\pi} d\phi
 \,\, \exp \{-i(N-\sum_i n_i) \phi \} \; ,
\end{equation}
Eq. (12) becomes:

\begin{equation}
 P(N) = \frac{1}{2\pi} \int_0^{2\pi} d\phi \,\, e^{-iN\phi}
 \prod_i
 \frac {1-e^{-\varepsilon_i/T}}{1-e^{-\varepsilon_i/T+i\phi}} \; .
\end{equation}
In the limit of a continuum of energy levels Eq. (14) can be
rewritten as:

\begin{equation}
 P(N) =  \frac{1}{2\pi} \int_0^{2\pi} d\phi \,\, e^{-iN\phi}
 \exp \{(2J+1) \frac{V}{(2\pi)^3} \int d^3 p \,\,
 \log {\frac {1-e^{-\varepsilon/T}} {1-e^{-\varepsilon/T+i\phi}}}\} \; .
\end{equation}
where $\varepsilon=\sqrt{p^2+m^2}$ and $J$ is the spin of
the particle.\\
Eq. (15) in this form can be transformed in an
integral over the unitary circle
in the complex plane. Let $w=\exp\{i\phi\}$ and

\begin{equation}
 I=\exp \{(2J+1) \frac{V}{(2\pi)^3} \int d^3 p \,\,
 \log \, (1-e^{-\varepsilon/T})\} \; .
\end{equation}
Then:

\begin{equation}
 P(N) = \frac{I}{2\pi i} \oint \frac{dw}{w^{N+1}} \exp \{-(2J+1)
 \frac{V}{(2\pi)^3} \int d^3 p \,\, \log \,
 (1-e^{-\varepsilon/T}w)\} \; .
\end{equation}
The residuals theorem can be used to get the final
expression of $P(N)$:

\begin{equation}
 P(N) = \frac {I}{N!} \, \lim_{w\rightarrow 0} \,\,
 \frac{d^N}{dw^N} \exp\{
  -(2J+1) \frac{V}{(2\pi)^3} \int d^3 p \,\, \log \,
  (1-e^{-\varepsilon/T}w)\} \; .
\end{equation}
It turns out that the derivatives in $w=0$ can be expressed
as a function of the functions $z_{(n)}$ defined as:

\begin{equation}
 z_{(n)} = \frac{V}{(2\pi)^3} \int d^3 p \,\,
 e^{-n\sqrt{p^2+m^2} /T} =
 (2J+1) \, \frac{VT}
 {2\pi^2 n} \, m^2 K_2 (\frac{nm}{T})
\end{equation}
where $K_2$ is the modified Bessel function of order 2.\\
Finally, by using the above expression of $P(N)$ one gets
the $\pi$-distribution
$f_\pi (n_1,n_2,n_3) = P_+(n_1)\, P_0(n_2)\, P_-(n_3)$.
In Table 4 we show probabilities $P_+(N)$ of producing
$N$ $\pi^+$
up to $N=6$ at $T=163$ MeV and $V=20$ Fm$^3$. The
probability values are compared
with those obtained from a Poisson distribution with
the same average value $\bar N$.

%

\newpage

\section*{Figure captions}

\begin{itemize}

\medskip

\item[\rm Figure 1]
Multiplicity distributions calculated in the framework
of the thermodynamical model at $\sqrt s = 91.2$ GeV
(dots) and corresponding Negative Binomial fits
(solid lines): a) charged tracks in full phase space,
b) charged tracks in a single hemisphere, c) primary hadrons.

\item[\rm Figure 2]
Multiplicity distributions calculated in the framework
of the thermodynamical model at $\sqrt s = 29 \div 35$
GeV (dots) and corresponding Negative Binomial fits
(solid lines): a) charged tracks in full phase space,
b) charged tracks in a single hemisphere, c) primary hadrons.

\item[\rm Figure 3]
Charged tracks multiplicity distributions in full
phase space; comparison between predictions of the
thermodynamical model (histogram) and data
(dots) \cite {del2,tasso} at a) $\sqrt s = 91.2$ GeV
and  b) $\sqrt s = 29 \div 35$ GeV.

\end{itemize}

\newpage

\section*{Tables}

\begin{table*}[h]
   \begin{center}
   \begin{tabular}{| c || c | c |}
      \hline
  {\bf Parameters}&$\sqrt s =91.2$ GeV &$\sqrt s = 29 \div 35$ GeV
  \\ \hline\hline
Temperature(MeV)  &  162.9$\pm$2.1   &  169.3$\pm$3.5  \\ \hline
Volume(Fm$^3$)    &  21.4$\pm$1.9    &  9.3$\pm$1.4    \\ \hline
$\gamma_s$        &  0.696$\pm$0.027 &  0.811$\pm$0.046  \\ \hline
$x_c$             &  0.884$\pm$0.029 &  0.89$\pm$0.12 \\ \hline
$x_b$          &  0.695$\pm$0.016   &  0.47$\pm$0.15   \\ \hline
   \end{tabular}
\caption[]{Values
of the parameters of the thermodynamical model \cite{td2}.}
\end{center}
\end{table*}

\begin{table*}[hb]
   \begin{center}
   \begin{tabular}{| c || c | c || c | c |}       \hline
   \multicolumn{5}{|c|}{ }                        \\
   \multicolumn{5}{|c|}{\large{Full phase space}} \\
   \multicolumn{5}{|c|}{ }                        \\ \hline
  & \multicolumn{2}{|c|}{$\sqrt s = 91.2$ GeV} &
  \multicolumn{2}{|c|}{$\sqrt s = 34$ GeV} \\ \hline
{\bf Parameters}& Calculated            & Measured
&   Calculated          &   Measured         \\ \hline
$\overline{n}$  & $21.45\pm0.093\pm0.58$& $20.71\pm0.77$
& $12.79\pm0.08\pm0.50$ & $13.59\pm0.46$  \\ \hline
$k$             & $28.34\pm1.43\pm1.41$ & $24.33\pm0.71$
& $13.03\pm0.58\pm0.66$ & $52.63\pm5.6$   \\ \hline
$D$             & $6.14\pm0.07\pm0.17$  & $6.28\pm0.43$
& $5.04\pm0.06\pm0.50$  & $4.14\pm0.39$   \\ \hline
$\chi^2 / dof$  & $1.81\pm0.60$         & $2.35$
& $10.9\pm5.53$         & $2.38$          \\ \hline\hline
   \multicolumn{5}{|c|}{ }                        \\
   \multicolumn{5}{|c|}{\large{Single hemisphere}}    \\
   \multicolumn{5}{|c|}{ }                        \\  \hline
   & \multicolumn{2}{|c|}{$\sqrt s =91.2$ GeV}
   & \multicolumn{2}{|c|}{$\sqrt s = 34$ GeV} \\ \hline
{\bf Parameters}& Calculated            & Measured
&   Calculated         &   Measured     \\ \hline
$\overline{n}$  & $10.72\pm0.045\pm0.29$& $10.35\pm0.47$
& $6.39\pm0.035\pm0.25$& $6.78\pm0.33$  \\ \hline
$k$             & $16.42\pm0.65\pm0.45$ & $15.06\pm0.39$
& $9.14\pm0.35\pm0.78$ & $50.0\pm5.0$   \\ \hline
$D$             & $4.21\pm0.035\pm0.056$& $4.19\pm0.32$
& $3.30\pm0.03\pm0.23$ & $2.78\pm0.22$  \\ \hline
$\chi^2 / dof$  & $1.76\pm0.30$         & $2.87$
& $8.83\pm2.8$         & $4.68$         \\ \hline
   \end{tabular}
\caption[]{Results of Negative Binomial fits to calculated
charged tracks multiplicity
distributions compared to results of analogous fits to
experimental data \cite{del1,tasso}.
The first quoted error is the fit error, the second one
is due to the uncertainty on parameters
of the thermodynamical model. The dispersion $D$ is
determined with the formula $D^2 =
\overline{n} + \overline{n}^2/k$.}
\end{center}
\end {table*}

\vspace{1cm}

\begin{table*}[h]
   \begin{center}
   \begin{tabular}{| c || c | c |}
      \hline
  {\bf }  &$\sqrt s =91.2$ GeV  &$\sqrt s = 29 \div 35$ GeV
  \\ \hline\hline
 \vspace{-0.35cm}
 & & \\
 $\overline{N}_c$ & $15.97\pm0.20\pm0.49$ & $8.91\pm0.12\pm0.78$
 \\ \hline
 \vspace{-0.35cm}
 & & \\
 $\overline{N}_p$ & $17.10\pm0.07\pm0.55$ & $9.39\pm0.05\pm0.36$
 \\ \hline\hline
 \vspace{-0.35cm}
 & & \\
 $k$              & $41.42\pm3.09\pm9.45$ & $32.18\pm3.11\pm26.0$
  \\ \hline
 \vspace{-0.35cm}
 & & \\
 $\chi^2 /dof$    & $1.63\pm0.60$   & $2.04\pm0.99$ \\ \hline
 \end{tabular}
\caption[]{Average number of clans $\overline{N}_c$
extracted from the Negative Binomial fits to charged
tracks multiplicity distributions compared with the
average number of primary hadrons $\overline{N}_p$.
Also shown are the results of a Negative Binomial fit
to the calculated primary
hadron distribution ($k$ and $\chi^2$). The first
quoted error is the fit error, the
second one is due to the uncertainty on parameters
of the thermodynamical model.}
\end{center}
\end{table*}

\begin{table*}[hb]
   \begin{center}
   \begin{tabular}{| c || c | c |}
      \hline
  {$N$}  &  $P_+(N)$  & Poisson    \\ \hline\hline
    0    &  0.311     & 0.293      \\ \hline
    1    &  0.346     & 0.359      \\ \hline
    2    &  0.208     & 0.221      \\ \hline
    3    &  0.0903    & 0.0904     \\ \hline
    4    &  0.0320    & 0.0278     \\ \hline
    5    &  0.00998   & 0.00682    \\ \hline
    6    &  0.00287   & 0.00139    \\ \hline
   \end{tabular}
\caption{Probability of producing $N$ $\pi^+$ at $T=163$ MeV
and $V=20$ Fm$^3$ compared to a Poisson distribution with the
same average value.}
\end{center}
\end{table*}
\clearpage

\newpage
\begin{center}
\begin{figure}[htbp]
\mbox{\epsfig{file=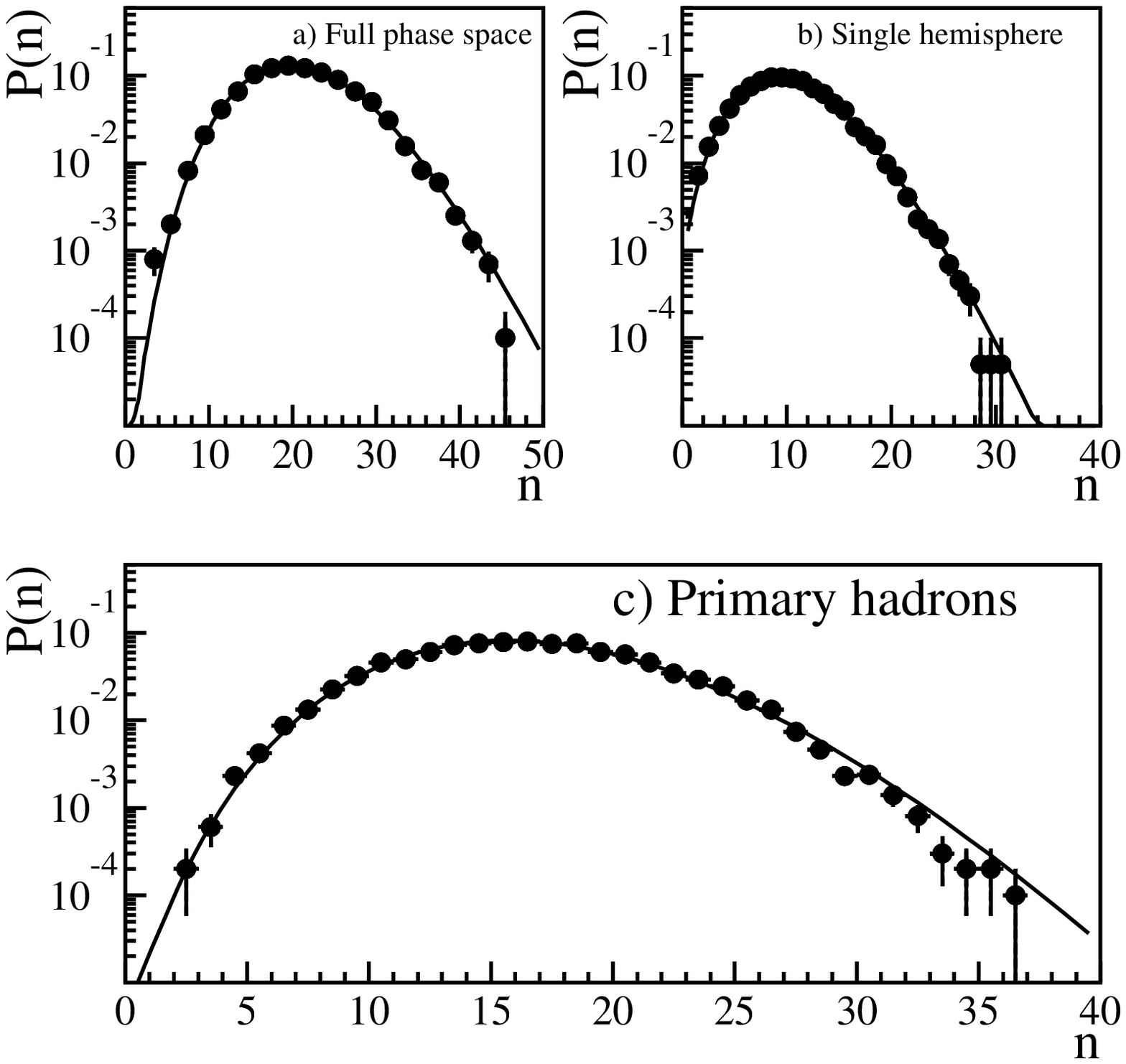,width=17cm}}
\caption{}
\end{figure}
\end{center}

\newpage

\begin{center}
\begin{figure}[htbp]
\mbox{\epsfig{file=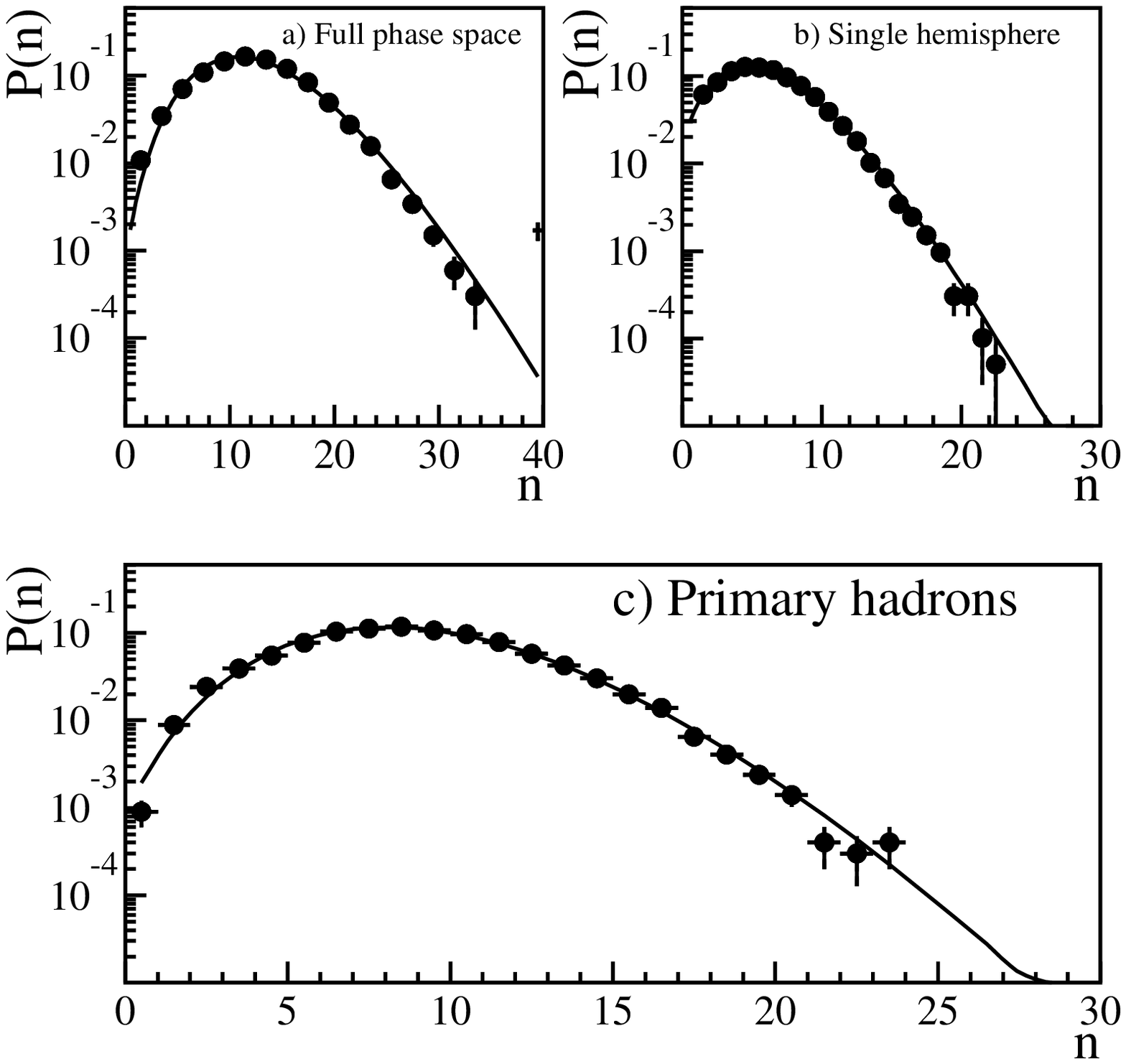,width=17cm}}
\caption{}
\end{figure}
\end{center}

\newpage

\begin{center}
\begin{figure}[htbp]
\mbox{\epsfig{file=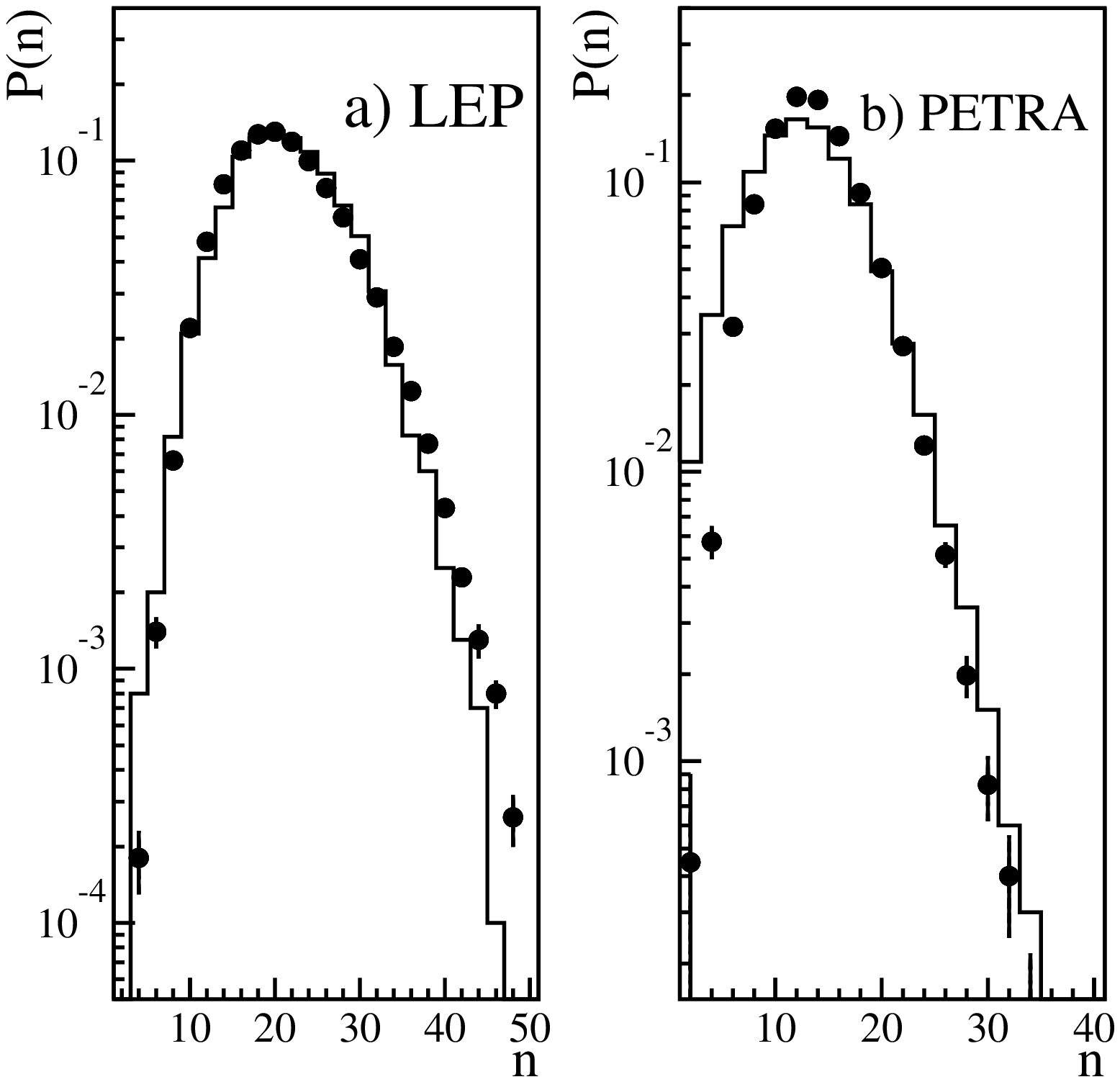,width=17cm}}
\caption{}
\end{figure}
\end{center}

\end{document}